\documentclass[class.elsarticle,aps,showpacs,twocolumn,floatfix,showkeys,10pt]{revtex4-1}
\usepackage{color}

\usepackage{colordvi}

\usepackage{graphicx,amssymb,epsfig,amsmath}
\usepackage{epstopdf}
\usepackage{multirow}

\newcommand{\be}{\begin{equation}}
\newcommand{\ee}{\end{equation}}
\newcommand{\bea}{\begin{eqnarray}}
\newcommand{\eea}{\end{eqnarray}}

\begin{document}
\title{Role of conditional shape invariance symmetry property to obtain eigen-spectrum of the generalized polynomial potential with a Coulomb term}
\author{ Sudesna Bera$^{1}$\footnote{E-mail addresses: sudesna251210@gmail.com (SB)  }, Rajesh Kumar Yadav$^{2}$\footnote{rajeshastrophysics@gmail.com (RKY)}, Barnali Chakrabarti$^{1}$\footnote{barnali.physics@presiuniv.ac.in (BC) }, Bhabani Prasad Mandal$^{3}$\footnote{bhabani.mandal@gmail.com (BPM)} }
\affiliation{
$^1$Department of Physics, Presidency University, 86/1 College Street, Kolkata 700083, India.\\
$^2$Department of Physics, S. P. College (S. K. M. University), Dumka-814101, India.\\
$^3$Department of Physics, Banaras Hindu University, Varanasi-221005, India.}

\begin{abstract}
A method based on supersymmteric (SUSY) quantum mechanics has been developed by exploiting conditional Shape invariance property 
for obtaining exact ground state solution of generalized polynomial potential with  Coulomb term. Specific cases have been discussed 
with extensive analytical calculation. How this method can be used to calculate the excited states has also been demonstrated. We have also
used a numerical technique (RKGS) and obtained the energy eigenvalues upto second excited state by solving 
the Schr\"odinger equation for quartic and sextic polynomial potentials with the Coulomb term and shown that the analytical results 
provide very good approximations to the numerical results. 
\end{abstract}

%\begin{keyword}
%SSQM, Conditional shape-invariance, Generalized Polynomial potential with the Coulomb term.
%\end{keyword}

\maketitle

\section{Introduction}
\hspace*{.5cm}

 One of the main focus of quntum mechanics is to find new solvable potentials. Supersymmetric quantum mechanics provides a very efficient and effective framework to find new solvable potential and it has been used extensively. Still there are some potentials which are still usolved. One such potential is polynomial potential with Coulomb term. It has been widely used in many problems. In this paper we are going to propose a indegenious method to solve generalized polynomial potential i.e. polynomial potential with any odd and even power term with a Coulomb term. 
The polynomial potential with a Coulomb term, which involves various powers of $r$, has physical significance in different branches 
of physics \cite{1,2,3,4,5,6}. This potential has been discussed and dissected by a number of researcher and with variety of techniques \cite{7,8,9,10,11,12}. It is already an established 
fact that exact ground state solutions of these potentials exist, provided certain conditions are satisfied between the potential parameters i.e. this potential belongs to the class of Conditionally Exactly Solvable potential class. Detailed study for 
specific cases are given and investigation on even power polynomial potentials has been also done \cite{13,14,15,16,17}.\\
The purpose of the revisit of the above problem is to utilize the shape invariance (SI) between the supersymmetric(SUSY) partner potentials and to establish
the fact that how the conditional shape invariance property leads to conditionally exact solution not only for ground state but also for higher excited state. It is shown by Gendenstine that whenever
the SI is satisfied by SUSY partner potential, the entire energy spectra including the eigenfunctions are determined
by algebraic means \cite{18} and this idea of SI has been extensively used for the exactly solvable potentials \cite{19}. But the
conditionally exactly solvable (CES) potentials belong to a very special class which cannot be tackled in general by Gendenshtein shape invariance prescription.
Thus in the present manuscript we study a very important class of CES potential i.e the generalized polynomial potential with a Coulomb term having both the 
even and odd power of $r$. To solve polynomial potential with a Coulomb term
we start with a superpotential ansatz and establish the SI condition between the partner potential of the Hamiltonian hierarchy. In this case we observe 
that for polynomial potential with a Coulomb term , the SI condition is satisfied only when the parameters in the potential satisfies some conditions and
we call this as `conditional shape-invariance.' Thus for all kinds of polynomial potentials as described below ground state of any member of Hamiltonian hierarchy can be calculated when the particular parametric constraint 
condition is satisfied. In this paper, after presenting the extensive calculation for the Quartic and Sextic polynomial potentials with a Coulomb term,  we also present the general $n$-term polynomial problem. For each case we show that the conditional SI  is the underlying symmetry to obtain the exact conditional ground state solutions for the polynomial
potential with a Coulomb term.\\
Along with proving CES symmetry as underlying reason for getting exact ground state solution of polynomial potential with Coulomb term, we also present a new and powerful methodology to get eigen spectrum of this kind of CES potential. To achieve that we extend the SUSY algebra by forming hierarchy of Hamiltonians. Now from SUSY formalism we know that $E_{n+1}^1=E_n^2$, so we always look for the ground state of next member of Hamiltonian  hierarchy. But as polynomial potential with a Coulomb term is a CES potential, calculation of ground state of every partner potential yields  different constraint condition which has to be satisfied
by potential parameters. So in case of CES potentials, to obtain the eigen-spectrum, the catch is the parametric constraint condition of potential parameters which has to be satisfied in every step of Hamiltonian hierarchy. \\ 
The paper is organized as follows. In section $II$ we give a brief overview of SUSY algebra. Section $III$-$IV$ deals with the specific problems of Quartic polynomial potential with a Coulomb term and Sextic polynomial potential with a Coulomb term. Section V deals with generalized polynomial potential which then followed by concluding remarks.
\begin{flushleft}
 \section{Supersymmetric algebra and the shape invariance}
\end{flushleft}
\hspace*{.5cm}
   To discuss the basic SUSY formalism,we start with the SUSY Hamiltonian in one dimension as  $ H= \left( \begin{array}{cc}
                                                               H_{1} & 0\\
                                                               0 & H_{2}\\
                                                              \end{array} \right)$, 
where $H_1$ and $H_2$ are two partner Hamiltonians, derived from factorizing $H$ and are defined as
\begin{equation}
 H_{1,2}= -\frac{d^{2}}{dx^{2}}+V_{\mp}(x)
\end{equation}
where the potential $V(x)$ is also factorized in two supersymmetric partner potential $V_{\mp}$.
These two partner  potentials are represented through the Riccati equation as 
\begin{equation}
 V_{\mp}= W^{2}\mp W^{\prime}(x)
\end{equation} 
where $ W(x)$ is the superpotential and is related to ground state wave function $\Psi_{0}(x)$ of $H_{1}$ by
\begin{equation}
 W(x)=-\frac{\Psi_{0}(x)^{\prime}}{\Psi_{0}(x)}.
\end{equation}
It is to be noted that energy level (in the units of $ \frac{\hbar}{\sqrt{2m}}$) is shifted to make the ground state energy of $H_1$ zero. Now in terms of superpotential, the generalized
annihilation and creation operator $A$ and $A^{\dagger}$ are defined as, 
\begin{eqnarray}
 A=\frac{d}{dx}+W(x), \quad A^{\dagger}=-\frac{d}{dx}+W(x).
\end{eqnarray}
 Thus the two partner Hamiltonians can be written as 
\begin{eqnarray}
 H_1= A^{\dagger}A, \quad  H_2= AA^{\dagger}. 
\end{eqnarray}
Using SUSY algebra, the correspondence between energy level of the two partner Hamiltonians can be shown easily. Let the eigenfunction of $H_{1,2}$ that corresponds
to eigenvalue $E^n_{1,2}$ be $\Psi^{n}_{(1,2)}$. Then the Schr\"odinger equation for $H_1$ is
\begin{eqnarray}
%\begin{split}
 H_1\Psi^{n}_{(1)}=A^{\dagger}A\Psi_{n}^{(1)}
                 =E_n^{(1)}\Psi_{n}^{(1)}
%\end{split}
\end{eqnarray}
which implies 
\begin{eqnarray}
\begin{split}
 H_2(A\Psi_{n}^{(1)})= AA^{\dagger}(A\Psi_{n}^{(1)}) 
                    = AH_1\Psi_{n}^{(1)}\\
                    = E_n^{(1)}(A\Psi_{n}^{(1)}). 
\end{split}
\end{eqnarray} 
Similarly the Schrödinger equation for $H_2$ is 
\begin{eqnarray}
 H_2\Psi_n^{(2)}= AA^{\dagger}\Psi_n^{(2)}
                 =E_n^{(2)}\Psi_n^{(2)}
\end{eqnarray}
which again implies 
\begin{eqnarray}
\begin{split}
 H_1(A^{\dagger}\Psi_n^{(2)})= A^{\dagger}A(A^{\dagger}\Psi_n^{(2)})
                       =A^{\dagger}H_2\Psi_n^{(2)}\\
                       = E_n^{(2)}(A^{\dagger}\Psi_n^{(2)}). 
\end{split}
\end{eqnarray}
Since $A\Psi_{0}^{(1)}=0$, ground state of $V_1(x)$ does not have a SUSY partner. So the pair of potential $V_{1,2}(x)$ have the same eigenspectra, only the ground state of $V_{1}(x)$ is missing in $V_{2}(x)$ i.e $E_{n+1}^{1}=E_{n}^{2}$.\\ 
 Gendenshtein first pointed out that the shape-invariance condition is mathematically expressed as 
\begin{equation}
 V_{2}(x,a_{1})=V_{1}(x,a_{2})+R(a_{1}),\quad   a_{2}= f(a_{1})
\end{equation}
where $a_{1}$ and $a_{2}$ are parameters appearing in the potentials and $f$ is a prescription for getting $a_2$ from $a_1$ and the remainder $R(a_1)$ is independent of $x$. To be similar in shape, the partner potentials must have same mathematical structure, but with altered parameters. 
The shape invariance condition is an integrability condition
and applying SUSY algebra for a hierarchy of Hamiltonian along with the shape invariance condition, one obtains the full eigen-spectrum of $H_{1}$ as 
\begin{equation}
 E_{n}^{1}= \sum_{k=1}^{n} R(a_{k})
\end{equation}
We will show that for CES potentials, the shape invariance criteria is satisfied only when a set of constraint condition involving potential parameters 
are satisfied and in this case also eigen states can be calculated by forming consecutive members of hierarchy of Hamiltonian.      
\section {Case I-Conditionally exact solution of Quartic Polynomial potential with a Coulomb term }
\hspace*{.5cm}
As a first case to demonstrate conditional shape invariance symmetry property and it's effective use to calculate eigen states, we start with polynomial potential with a Coulomb term up to 4th power in $r$. So the form of this particular potential is
\begin{equation}\label{1}
 V(r)=ar+br^{2}+dr^{3}+fr^{4}+\frac{c}{r}+\frac{l(l+1)}{r^{2}}; \quad f>0.
\end{equation}
Now to calculate the ground state using SUSY formalism, we take $ V_1(r)=V(r)-E_0$ , where $E_0$ is the ground state energy of $V(r)$. 
We start with the following superpotential anstaz
\begin{equation}\label{4}
 W(r)=Ar+A_{1}r^{2}-\frac{B}{r}+D; \quad B>0, D>0,
\end{equation}
where $W(r)$ is related with $V_1(r)$ as
\begin{eqnarray}\label{2}
 V_1(r)= V(r)-E_0= W^2(r)-W^{\prime}(r).
\end{eqnarray}
From Eqns. (\ref{1}) and (\ref{2}), comparing the coefficients of same power of $r$, the unknown parameters $A$, $A_1$, $B$ and $D$ are calculated. These unknown parameters are related to the known potential parameters $a,b,c,d$ and $f$ which satisfy the following relations
\begin{eqnarray}
\begin{split}
 A_1^2=f,\quad 2AA_1=d,\quad B^2-B=l(l+1),\nonumber\\
 -2BD= c,\quad A^2+2A_1D=b,\quad 2AD-2A_1B-2A_1= a,
\end{split}
 \end{eqnarray} 
\begin{equation}\label{3}
  \mbox{and}\qquad  -E_0= D^2-2AB-A.
\end{equation}
The solutions of the above equations are 
\begin{eqnarray}
B&=&l+1, \qquad A_1=\sqrt{f},\qquad A=\frac{d}{2\sqrt{f}}\nonumber\\
\mbox{and} \qquad D&=&-\frac{c}{2(l+1)}.
\end{eqnarray}
Using these parameters in Eq.(\ref{3}), the ground state energy is calculated as
\begin{equation}
 E_{0}=-\frac{c^{2}}{4(l+1)^{2}}+\frac{d(l+1)}{\sqrt{f}}+\frac{d}{2\sqrt{f}}.
\end{equation}
The validity of this equation is subjected to the satisfaction of the constraint conditions involving potential parameters, which are given by
\begin{equation}\label{5}
 a=-\frac{cd}{2\sqrt{f}(l+1)}-(2l+4)\sqrt{f} , \qquad 
 b=\frac{d^{2}}{4f}-\frac{c\sqrt{f}}{(l+1)}.
\end{equation}
The ground state wave function can be easily calculated by using $\hat{A}\psi_{0}(r)=0$, given as
\begin{equation}
\begin{split}
 \psi_{0}^{(f,d,c)}(r)=N_{0}r^{l+1}\exp\bigg[\bigg(\frac{c}{2(l+1)}\bigg)r\\
-\bigg(\frac{d}{4\sqrt{f}}\bigg)r^{2}-\bigg(\frac{\sqrt{f}}{3}\bigg)r^{3}\bigg].
\end{split}
\end{equation}
Now to demonstrate the conditional SI, we construct the SUSY partner potential $V_{2}$ as
\begin{eqnarray}
\begin{split}
 V_2(r)=W^2(r)+W^{\prime}(r)\\
       =a_1r+b_1r^{2}+d_1r^{3}+f_1r^{4}+\frac{c_1}{r}+\frac{l_1(l_1+1)}{r^{2}}
\end{split}
\end{eqnarray}
Using $W(r)$ and comparing the coefficients of $r$, we see that three potential parameters ($d,f,c$) remain same while other two parameters are changed from the original potential parameter. Actually they are translated and related with the old parameters through the equation
\begin{eqnarray}
 a_{1}= 2AD-2A_1B+2A_1 = a+4\sqrt{f}
\end{eqnarray}
\begin{eqnarray}
 l_{1}(l_{1}+1)= B^2+B = B^2-B+2B= (l+1)(l+2).
\end{eqnarray}
And there is a extra constant term in $V_2(r)$ which is given by 
\begin{equation}
 R= \frac{c^{2}}{4(l+1)^{2}}-\frac{d(l+1)}{\sqrt{f}}+\frac{d}{2\sqrt{f}}.
\end{equation}
 So from the above analysis, we observe that $V_2(r)$ is basically shape invariant with the translation of parameters $l \rightarrow  l_1  \rightarrow l+1$
and  $a\rightarrow a_1\rightarrow a+4\sqrt{f}$.\\
 But this shape invariance is subjected to the fulfilment of the abovesaid constraint conditions between the potential parameters. Therefore the 
quartic polynomial potential with a Coulomb term represents an example of conditional SI symmetry. It is obvious from the disscusion above that $V_2(r)$ has different potential parameters than $V_1(r)$ as some of the parameters of $V_2(r)$ has been translated. It is to be noted that only the parameters calculated from constraint conditions have been changed while the independent potential parameters $c$, $d$ and $f$ have remained same. 
 Now we are claiming that this conditional SI property is the underlying symmetry to get the exact ground state of CES potentials.\\
To calculate the ground state of $V_2(r)$ we carry forward the present prescription and construct the potential $V_2^{\prime}(r)$ as 
\begin{equation}
 V_2^{\prime}(r)=V_2(r)-E_0^{(1)},
\end{equation}dissected
where $E_0^{(1)}$, the ground state energy of $V_2(r)$ in the shifted energy scale, is subtracted to make the ground state energy of $V_2(r)$ zero. 
\begin{equation}
 V_2^{\prime}(r)=a_1 r+b_1 r^{2}+d r^{3}+f r^{4}+\frac{c}{r}+\frac{l_1(l_1+1)}{r^{2}}+R+E_0-E_0^1.
\end{equation}
Since $V_2(r)$ has the same $r$ dependence as that of  the $V_1(r)$ thus the form of the superpotential $W_1(r)$ of $V_2(r)$ will be same as 
given in Eq.(\ref{4}). Repeating the same calculation as before, we find the constraint conditions
for the ground state of $V_2(r)$ as
\begin{equation}
 a_{1}=-\frac{cd}{2\sqrt{f}(l_1+1)}-2(l_1+1)\sqrt{f}-2\sqrt{f},
\end{equation}
\begin{equation}
 b_{1}=\frac{d^{2}}{4f}-\frac{c\sqrt{f}}{(l_1+1)}
\end{equation}
and the ground state energy of $V_2^{\prime}(r)$ in the shifted energy scale becomes
\begin{equation}
 E_0^{1}= -\frac{c^{2}}{(l_1+1)^{2}}+\frac{d(l_1+1)}{\sqrt{f}}+\frac{d}{2\sqrt{f}}+R.
\end{equation}
Now as we have calculated the ground state of $V_2^{\prime}(r)$ by making the the ground state energy of $V_2(r)$ zero as required by SUSY,
 thus the calculated energy have to be shifted properly. So in the proper energy scale, the ground state energy of $V_2(r)$ is given by
\begin{eqnarray}
 E_0^{V_2}= E_0^1+E_0^{V_1}
          =-\frac{c^2}{4(l_1+1)^2}+\frac{d(l_1+1)}{\sqrt{f}}+\frac{3d}{2\sqrt{f}}.
\end{eqnarray}
where $E_0^{V_1}$ is actually ground state energy of $V(r)$ i.e. $E_0^{V_1}\equiv E_0$ and $E_0^1$ is the ground state energy of $V_2(r)$ in shifted energy scale. 
\subsection{Observation}
It is evident that by following the same procedure, we can construct the partner potential of $V_2$ (say $V_3$), 
the partner potential of $V_3$ (say $V_4$) and so on. 
From the previous calculations it has been also observed that every consecutive member of hierarchy of Hamiltonian is SI
 under certain constraint conditions on the potential parameters. It has been shown that
the constraint conditions will change in each step in the hierarchy. For example, the parameter $l$ is translated to $l+1$ $\rightarrow$ $l+2$ in each step in the hierarchy of Hamiltonian.
So, the quartic polynomial potential with a Coulomb term is a CES potential as it satisfies SI  property subjected to 
certain constraint conditions involving potential parameters.\\
This concept of conditional SI symmetry can be extended to calculate the excited states of CES potentials. 
For a truly exactly solvable SI potentials, the ground
states of $n$th member of hierarchy is basically the $(n-1)$-th excited state of original stationary Hamiltonian. 
In the case of CES potentials also, the method of calculating excited states by forming hierarchy of Hamiltonian can be extended. 
But in every consecutive steps there will be two new constraint conditions involving potential parameters of the member of hierarchy which means potential parameters of the partner potential will also change. As the potential parameters of partner potentials change in each step, the accuracy of this methodology highly depends on 
the judicious choice of parameters.\\
So besides showing CES symmetry of quartic polynomial potential with a Coulomb term, the new finding of our works is the degeneracy property of the member of hierarchy of Hamiltonian i.e
\begin{equation}
 E_1^{V_1}= E_0^{V_2},
\end{equation}
\begin{equation}
 E_2^{V_1}= E_1^{V_2}=E_0^{V_3}.
\end{equation}
%Note that in each step different constraint condition is satisfied and the parameter $l$ is translated to $l+1$ and so on.\\
 The above results are generalized to calculate analytically the ground state energy of the $n$th member of hierarchy of Hamiltonian along with constraint conditions in terms of the independent potential parameters of the original potential. 
The expression for energy is given by
\begin{equation}
 E_0^{V_n}= -\frac{c^2}{4(l+n+1)^2}+\frac{d(l+n+1)}{\sqrt{f}}+\frac{d(1+2n)}{2\sqrt{f}},
\end{equation}
with the constraint conditions  
\begin{equation}
 a_n= \frac{cd}{2\sqrt{f}(l+n+1)}-2(l+n+1)\sqrt{f}-2\sqrt{f}
\end{equation} 
and
\begin{equation}
 b_n=\frac{d^2}{4f}-\frac{c\sqrt{f}}{l+n+1}.
\end{equation}
As it is obvious from the previous calculation, $l$, $d$, $f$, $c$ are independent parameters and $a$ and $b$ are dependent parameters, as they are calculated from the given constraint condition. So starting with these independent parameters, one can calculate
ground state of any member in the hierarchy of Hamiltonian. But as the potential parameters changes, this method will be successful only for a restricted set of parameters.
As our finding is general, it is true for other polynomial class conditionally shape invariant potentials also.
\subsection{Result} 
In this section, we present analytically calculated energy and compare it with the numerically calculated solution of Schrödinger equation involving quartic polynomial potential with a Coulomb term.
For different choices of $c$, $d$, $l$ and $f$, we have calculated the parameters $a$ and $b$ from the above constraint conditions Eq.(\ref{5}). 
For the first choice (SET I) we have taken $l=1$, $d=0.5$, $c=-1.0$ and $f=0.1$. The value of $a$ and $b$ is given in the TABLE I. In the above said table, we have presented the result upto 4th decimal place.  Here we have calculated energy upto the second excited states and compared our result with the numerical solutions of the
Schrödinger equation by using RKGS (Range-Kuuta) method. The results are tabulated below in TABLE II and energy values are presented upto 2nd decimal place.   

\begin{table}[h]
\centering
\caption{Caculated value of potential parameters $a$, $b$ from constraint conditions for two set of parameters.}
\label{my-label}
\begin{tabular}{|c|c|c|c|c|}
\hline
\multicolumn{4}{|c|}{Dependent Potential Parameters} \\
\hline
 SET I   & a=$-$1.5020     & a$_1$=$-$2.2662   &a$_2$= $-$2.9646 \\  \cline{2-4} 
            &b=.7831       &b$_1$=.7304        & b$_2$= 0.7040             \\ \hline
 SET II    & a=$-$1.559      & a$_1$=$-$2.0977 &a$_2$=$-$2.6315           \\ \cline{2-4}
            &b=0.3346         &b$_1$=0.3302    & b$_2$= 0.3280                  \\   \hline              
\end{tabular}
\end{table}

Similar to the first set, for the second choice (SET II in TABLE I) we have taken $l=1$, $d=0.3$, $f=0.07$, $c=-0.1$ and calculated $a$ and $b$. The analytically calculated energy values (upto second excited states) are compared with the numerical results and are listed in TABLE II. 
\begin{table}[h]
\centering
\caption{Comparison of ground state and excited state energy for two set of potential parameters}
\label{my-label}
\begin{tabular}{|c|c|c|c|c|c|c|}
\hline
\multicolumn{6}{|c|}{Energy calculation for two set of parameters} \\
\hline
\multicolumn{3}{|c|}{SET I} 
&
\multicolumn{3}{|c|}{SET II} \\
\hline
V$_1$                             & V$_2$           &V$_3$               & V$_1$                  & V$_2$                   &V$_3$            \\ \hline
E$^{S}_{0}$=3.89              & $-$               &$-$               &E$^{S}_{0}$=2.83               & $-$               &$-$            \\ \cline{1-1} \cline{4-4}
E$^{R}_{0}$=3.89               &                    &                  &E$^{R}_{0}$=2.83                &                 &              \\ \hline
                                                         
E$^{R}_{1}$=7.06           & E$^{S}_{0}$=7.09   &$-$                & E$^{R}_{1}$=5.06        & E$^{S}_{0}$=5.10    &$-$    \\ \cline{2-2} \cline{5-5}
                                &E$^{R}_{0}$= 7.08   &                   &                          & E$^{R}_{0}$= 5.09     &     \\ \hline
E$^{R}_{2}$=10.15             &  E$^{R}_{1}$= 10.23   & E$^{S}_{0}$=10.26 & E$^{R}_{2}$=7.34 & E$^{R}_{1}$=7.35       & E$^{S}_{0}$= 7.37         \\ \cline{3-3} \cline{6-6}  
                                      &                      & E$^{R}_{0}$=10.24           &                         &          & E$^{R}_{0}$= 7.36         \\ \hline
\end{tabular}
\end{table}

 Here it is worth mentioning that the accuracy of our method highly depends on the choice of original potential parameters.  
If the independent parameters ($d$, $c$, $f$) are chosen in such a way that the difference between $a$ and $b$ of two consecutive partner potential is large, then the numerically calculated values and analytically calculated
values obtained from SUSY vary considerably. So by using the aforesaid method, excited states of Quartic polynomial potential with a Coulomb term can be calculated for a
restricted set of potential parameters. Accuracy/usefullness of this method mainly depend of two things, first: choice original potential parameters, second: which excited state we are looking for. We can say from our finding that for a trully  high excited state, the difference between analytically and numerically calculated energy values will be significant. 

\section{Case II-Conditionally exact solution of Sextic polynomial potential with Coulomb term }
\hspace*{.5cm}
In this section we consider the Coulomb potential perturbed by Sextic polynomial potential and solve by using the 
concept of conditional SI property. 
In this case the polynomial potential has terms up to $6$th order of $r$.
The form of the potential is 
\begin{equation}
 V(r)=\frac{l(l+1)}{r^{2}}+\frac{c}{r}+ar+br^{2}+dr^{3}+fr^{4}+g{r^{5}}+h{r^{6}}.
 \end{equation}
For this potential, the considered superpotential ansatz is
\begin{equation}
 W(r)=Ar+A_{1}r^{2}+A_{2}r^{3}-\frac{B}{r}+D.
\end{equation}
Following the same procedure as in the previous case, we find the ground state energy of $V(r)$ is
\begin{equation}
 E_{0}=-\frac{c^{2}}{4(l+1)^{2}}+2(l+1)[\frac{f}{2\sqrt{g}}- \frac{g^{2}}{8(h)^{\frac{3}{2}}}]+[\frac{f}{2\sqrt{g}}-\frac{g^{2}}{8(h)^{\frac{3}{2}}}],
\end{equation}
 with the following constraint conditions on the potential parameters 
\begin{eqnarray}
 a&=&-\frac{c}{(l+1)}(\frac{f}{2\sqrt{g}}- \frac{g^{2}}{8(h)^{\frac{3}{2}}})-(l+2)(\frac{f}{\sqrt{g}}),\nonumber\\
 b&=&(\frac{f}{2\sqrt{g}}- \frac{g^{2}}{8(h)^{\frac{3}{2}}})^{2}-\frac{cg}{2(l+1)\sqrt{h}}-(2l+5)\sqrt{h}\nonumber\\
\mbox{and}\qquad
 d&=&-\frac{c\sqrt{h}}{(l+1)}+\frac{fg}{2h}-\frac{g^{3}}{8h^{2}}.
\end{eqnarray}
The partner potential can be obtained easily and find that the three dependent
potential parameters $(a_{1}, b_{1},d_{1})$ are related with $(a, b, d)$ by the following relations
\begin{eqnarray}
 a_{1}=a+\frac{2f}{\sqrt{h}}, \qquad b_1=b+6\sqrt{h}, \qquad l_{1}=l+1,
\end{eqnarray}
which  shows the conditional SI between the partner Sextic polynomial potentials with Coulomb term. 
So indeed $V(r)$ is again a SI potential with translation of parameters but the shape-invariance property
is subjected to the fulfilment of the constraint condition involving potential parameters.
The ground state wave function is given as
\begin{equation}
\begin{split}
 \psi^{0}_{(c,f,g,h)}=N_{0}^{2}r^{l+1} \exp\bigg[-\big(\frac{c}{2(l+1}\big)r-\big(\frac{f}{2\sqrt{g}}\\
 - \frac{g^{2}}{8(h)^{\frac{3}{2}}}\big)\frac{r^{2}}{2}-\big(\frac{g}{\sqrt{h}}\frac{r^{3}}{6}\big)-\big(\sqrt{h}\frac{r^{4}}{4}\big)\bigg].
 \end{split}
\end{equation}
Now as mentioned earlier, we calculate the first excited state energy by constructing hierarchy of Hamiltonians.
The first excited state energy is thus given by 
\begin{equation}
 E_{1}=-\frac{c^{2}}{4(l_1+1)^{2}}+2(l_1+1)[\frac{f}{2\sqrt{g}}- \frac{g^{2}}{8(h)^{\frac{3}{2}}}]+3[\frac{f}{2\sqrt{g}}-\frac{g^{2}}{8(h)^{\frac{3}{2}}}],
\end{equation}
with the new constraint conditions 
\begin{eqnarray}
 a_{1}&=& -\frac{c}{(l_1+1)}(\frac{f}{2\sqrt{g}}- \frac{g^{2}}{8(h)^{\frac{3}{2}}})-(l_1+2)(\frac{f}{\sqrt{g}}),\nonumber\\
 b&=&(\frac{f}{2\sqrt{g}}- \frac{g^{2}}{8(h)^{\frac{3}{2}}})^{2}-\frac{cg}{2(l_1+1)\sqrt{h}}-(2l_1+5)\sqrt{h}\nonumber\\
 \mbox{and}\qquad d_{1}&=&-\frac{c\sqrt{h}}{(l_1+1)}+\frac{fg}{2h}-\frac{g^{3}}{8h^{2}}.
\end{eqnarray}
In this case also, the higher members of hierarchy can be calculated by the same procedure as discussed in the previous section.\\
Thus the sextic polynomial potential with a Coulomb term is also a conditionally SI potential and by exploiting it's conditional SI
 symmetry property the excited states can be calculated.
The general analytical expression for the ground state energy of $n$-th member of the hierarchy in terms of the independent parameters
 along with the constrain conditions are also obtained which are given as
\begin{equation}
\begin{split}
E_n= -\frac{c^2}{4(l+n+1)^2}+2(l+n+1)\bigg[\frac{f}{2\sqrt{g}}- \frac{g^{2}}{8(h)^{\frac{3}{2}}}\bigg ]\\
+(1+2n)\bigg[\frac{f}{2\sqrt{g}}- \frac{g^{2}}{8(h)^{\frac{3}{2}}}\bigg],\nonumber\\
\end{split}
\end{equation}
\begin{eqnarray}
%\begin{split}
 a_n&=&-\frac{c^2}{(l+n+1)^2}(\frac{f}{2\sqrt{g}}- \frac{g^{2}}{8(h)^{\frac{3}{2}}})-(l+n+2)(\frac{f}{\sqrt{g}}),\nonumber\\
 b_n&=&(\frac{f}{2\sqrt{g}}- \frac{g^{2}}{8(h)^{\frac{3}{2}}})^{2}-\frac{cg}{2(l+n+1)\sqrt{h}}-(2(l+n)+5)\sqrt{h},\nonumber\\
\mbox{and} \
 d_n&=&-\frac{c\sqrt{h}}{(l+n+1)}+\frac{fg}{2h}-\frac{g^{3}}{8h^{2}}.
%\end{split}
\end{eqnarray}
In this case there are three dependent potential parameters which has to be calculated from the constraint conditions. So for Sextic polynomial potential with a Coulomb term, the accurate result for the excited states energy 
can be produced for a more restricted set of parameters.

\section{Generalized Polynomial Potential with Coulomb term}
\hspace*{.5cm}
From the above two examples we observe that for different powers of polynomial potentials the form of superpotential, energy and the  
constraint conditions are consistent. In this section we consider a more general form of polynomial type potential upto $2n-th$ degree, 
where $n$ is any positive integer number and solve this potential by using the concept of conditional SI symmetry property.
Though this type of potential has already been discussed before \cite{20}, our main aim is to show that for a generalized case also, 
conditional shape invariance property is satisfied. 
The form of generalized polynomial potential with the Coulomb term  is given by
\begin{equation}
 V_{2n}(r)=\frac{l(l+1)}{r^{2}}+\frac{c}{r}+\sum_{i=1}^{2n}a_{i}r^{i},
\end{equation}
where $n$ is any positive integer number. To solve this potential problem using SUSY, we take the superpotential anstaz as 
\begin{equation}
 W_{n}(r)=-\frac{B}{r}+\sum_{i=1}^{n}A_{i}r^{i}+D.
\end{equation}
The ground state of $V_{2n}(r)$ is calculated by using the Riccati equation 
\begin{eqnarray}
 V_{2n}^{1}(r)&=&V_{2n}(r)-E_{0}=W^{2}(r)-W^{\prime}(r),\nonumber\\
 &=&D^{2}-\frac{2BD}{r}+\frac{B^{2}-B}{r^{2}}+\sum_{i,j=1}^{n}A_{i}A_{j}r^{i+j}\nonumber\\
&+&2D\sum_{i=1}^{n} A_{i}r^{i}-2B\sum_{i=1}^{n}A_{i}r^{i-1} -\sum_{i=1}^{n} (i)A_{i}r^{i-1},
\end{eqnarray}
where $E_0$ is the ground state energy of $V_{2n}(r)$ and  $i$ runs from $1$ to $n$. 
Now comparing the coefficients of same powers of $r$, we get $n$ number of constraint conditions for potential parameters and $n$ number of equations
 for superpotential parameters, which are given below
\begin{eqnarray}\label{12}
 B^2-B=l(l+1),\qquad -2BD=c,
\end{eqnarray}
and
\begin{eqnarray}
 a_{2n}&=&A_{n}^{2},\qquad a_{2n-1}=2A_{n-1}A_{n},\nonumber\\
 a_{2n-2}&=&A^{2}_{n-1}+2A_{n-2}A_{n},
\end{eqnarray}
or in general
\begin{equation}\label{7}
 (a_{k})_{odd}=\sum_{i,j=1}^{n}A_{i}A_{j}\quad i+j=k, i\neq j,
\end{equation}
where the subscript `odd' denotes that $'k'$ is a odd number which actually means the concerned potential parameter is a coefficient of $r$ with odd power term in the potential.
\begin{equation}\label{8}
 (a_{k})_{even}=\sum_{i,j=1}^{n}A_{i}A_{j}+A^{2}_{k/2} \quad i+j=k,  i\neq j,
\end{equation}
here also the subscript `even' denotes that $k$ is even, which in turn means that the potential parameter is a coefficient of even power of $r$. 
 For both the equations, 
$k $ runs from $ n+1 \rightarrow 2n $, where $n$ is any positive integer number. \\
In the similar way the $n$ number of constraint conditions are given as
\begin{equation}
\begin{split}
 (a_{m})_{odd}=2DA_{m}-2BA_{m+1}-(m+1)A_{m+1}\\
 +\sum_{i,j=1}^{n}A_{i}A_{j}, i+j=m,i\neq j
 \end{split}
\end{equation}
\begin{equation}
\begin{split}
 (a_{m})_{even}=A^{2}_{m-1}+2DA_{m}-2BA_{m+1}-(m+1)A_{m+1}\\
 +\sum_{i,j=1}^{n}A_{i}A_{j}, i+j=m,i\neq j
 \end{split}
\end{equation} where $m$ runs from $1 \rightarrow n $. Here also the subscript `odd', `even' has the same meaning as stated before. 
 If $n$ is even then
\begin{equation}
 a_{n}=A^{2}_{n-1}+2DA_n+\sum_{i,j}^{n}A_{i}A_{j} , i+j=n ,i\neq j, 
\end{equation}
and if $n$ is odd then
\begin{equation}
 a_{n}=2DA_{n}+\sum_{i,j}^{n}A_{i}A_{j} , i+j=n ,i\neq j. 
\end{equation}
After solving the Riccati equation for the ground state, we observe that $n$ number of potential parameters are independent i.e., 
they can be chosen freely while $n$ number of parameters are dependent as there are total $n$ number of 
constraint conditions involving potential parameters. 
The ground state energy can be calculated from the equation given below
\begin{equation}
 E_0= -D^2+2A_1B+A_1,
\end{equation}
where the super potential parameters are $B=l+1$, $D=-\frac{c}{2(l+1)}$ and $A$ can be calculated from Eq.(\ref{7}) and Eq.(\ref{8}).
Now to check whether SI condition is satisfied or not, we calculate the partner potential from the Riccati equation as
\begin{eqnarray}
V_{2n}^2(r)&=& W^{2}+W^{\prime}\nonumber\\
&=& D^{2}-\frac{2BD}{r}+\frac{B^{2}+B}{r^{2}}+\sum_{i,j}^{n}A_{i}A_{j}r^{i+j}+2D\sum_{i=1}^{n} A_{i}r^{i}\nonumber\\
&-&2B\sum_{i=1}^{n} A_{i}r^{i-1}+\sum_{i=1}^{n} (i)A_{i}r^{i-1}. 
\end{eqnarray}
Comparing coefficients of same powers of $r$, we find that
\begin{equation}\label{9}
 l^{1}= l+1.
\end{equation}
The independent potential parameters remain same, whereas in case of dependent $n$ numbers of parameters, 
partner potentials parameters are related with original potential parameters by the following equations.
\begin{equation}\label{10}
 (a_{m})_{odd}^{1}= (a_{m})_{odd}+ 2(m+1)A_{m+1}
\end{equation}
\begin{equation}\label{11}
 (a_{m})_{even}^{1}=(a_{m})_{even}+ 2(m+1)A_{m+1}
\end{equation}
where $m$ runs from $ 1 \rightarrow n$. 
 So, in the case of $V_{n}^{2}$ i.e the partner potential, the dependent potential parameters are related with the original potential parameters through Eq. (\ref{9}), Eq. (\ref{10}) and Eq. (\ref{11}). Thus the generalized polynomial potential with a Coulomb term is a SI potential with respect to translation of parameters. But as the potential parameter has to obey certain constraint condition
 to satisfy the shape invariance criteria, this is a case of conditional shape-invariance symmetry.\\
 And in this case also higher energy states can be calculated by forming hierarchy of Hamiltonian in the same way prescribed before for Quartic polynomial potential with Coulomb term. But as the order of polynomial increases, the task of calculating higher energy states will be more difficult and cumbersome. Also with increasing order of polynomial potential, there 
 will be more constraint condition on potential parameters which will reduce the accuracy of this method and set of suitable potential parameters will also be much more restricted. \\
 But using the generalized formula for superpotential and constraint conditions, any polynomial potential 
with Coulomb term can be solved. 
Here we have discussed two special cases for $n=1$ and $n=2$ briefly.\\
 The form of the Quadratic polynomial potential with Coulomb term is
 \begin{equation}
  V_2(r)=\frac{l(l+1)}{r^2}+\frac{c}{r}+\sum_{i=1}^2a_ir^i
        =\frac{l(l+1)}{r^2}+\frac{c}{r}+a_1r+a_2r^2
 \end{equation}
The form of the superpotential is 
\begin{equation}
 W_1(r)=-\frac{B}{r}+A_1r+D
\end{equation}
Here as $n=1$, $l$ is $1$ and $k$ is $2$. So there is one constraint condition involving the coefficient of $r$. The superpotential parameters are calculated by using Eqs.(\ref{12}), (\ref{7})  and (\ref{8}). So $A_1$ is calculated from
\begin{equation}
 a_2=A_1^2,
\end{equation}
which gives us $A_1=\sqrt{a_1}$, $B=l+1$ and $D=-\frac{c}{2(l+1)}$.
Using the generalized equation, Eq. (53) for n=1, we obtain the constraint conditions as 
\begin{equation}
 a_1=2DA_1=-\frac{c\sqrt{a_2}}{l+1}.
\end{equation}
Here as $m=1$, there is no other terms in the equation as they involve $(m+1)$ and $i, j$ are positive integer. So using the generalized expression we have calculated
all the superpotential parameters as well as the constraint condition. 
The ground state energy of $V_2(r)$ is
\begin{equation}
 E=-D^2+2A_1B+A_1=-\frac{c^2}{4(l+1)^2}+2(l+1)\sqrt{a_2}+\sqrt{a_2},
\end{equation}
 and for the partner potential of $V_2(r)$, the shape invariance criteria is given by 
 \begin{eqnarray}
  l^1=l+1,   \quad a_1^1=a_1=-\frac{c\sqrt{a_2}}{l_1+1}.
 \end{eqnarray}
 For $n=2$ case, form of the potential is
 \begin{equation}
 \begin{split}
  V_4(r)= \frac{l(l+1)}{r^2}+\frac{c}{r}+\sum_{i=1}^4a_ir^i \\ 
       =\frac{l(l+1)}{r^2}+\frac{c}{r}+a_1r+a_2r^2+a_3r^3+a_4r^4,
\end{split}
 \end{equation}
So the superpotential is given by as \begin{equation}
                                      W_2(r)=-\frac{B}{r}+A_1r+A_2r^2+D.
                                     \end{equation}
As here $n=2$, $m=1,2$, $k=3,4$. So there will be two constraint conditions. And besides calculating $B=l+1$, $D=-\frac{c}{2(l+1)}$, $A_1$ and $A_2$ will be calculated from the equation given below. 
So, to calculate the superpotential parameters $A_1$, $A_2$, using the generalized expression, the required equations are
\begin{eqnarray}
 a_4=A_2^2,  \quad
 a_3= 2A_1A_2
\end{eqnarray}
So, the solution for the superpotential parameters are $A_2=\sqrt{a_4}$, $A_1=\frac{a_3}{2\sqrt{a_4}}$. Using the generalized equation, the first constraint condition is
\begin{equation}
 a_1=2DA_1-2BA_2-2A_2=-\frac{c}{(l+1)}\frac{a_3}{2\sqrt{a_4}}-2(l+1)\sqrt{a_4}-2\sqrt{a_4}. 
\end{equation}
Here the last term has vanished because $i$ and $j$ are positive integer. 
In the similar way, deriving from Eq.(55) the second constraint condition is given by as 
\begin{equation}
 a_2=A_1^2+2DA_2= \frac{a_3^2}{4a_4}-\frac{ca_4}{(l+1)}.
\end{equation}
Thus the analytical expression for the ground state energy is 
\begin{equation}
 E_0=-D^2+2A_1B+A_1=-\frac{c^2}{4(l+1)^2}+(l+1)\frac{a_3}{\sqrt{a_4}}+\frac{a_3}{2\sqrt{a_4}}.
\end{equation}
If we replace $a_1=a$, $a_2=b$, $a_3=d$, $a_4=f$ then all the results derived in Sec. III are reproduced here very easily using the generalized expressions for calculation of superpotential parameters as well as constraint conditions. 
Now to check the conditional shape invariance criteria, from the generalized expression, we find that
\begin{eqnarray}
 a_1^1=a_1+4\sqrt{a_4}, \quad  l^1=l+1.
\end{eqnarray}
So from the above two example discussed, it is obvious that from the generalized expressions for superpotential parameters as well as constraint conditions, any polynomial potential with Coulomb term can be solved 
and exact ground state can be very easily calculated. 

\section{Conclusion}
\hspace*{.5cm}
In this paper, we have discussed extensively about the conditional SI symmetry of any order of polynomial potential with a Coulomb term. We have explicitly calculated the quartic and sextic polynomial potentials with a Coulomb term and obtained the ground as well as the excited energy states 
by using the idea of conditional SI symmetry. General analytical formula for calculation of energy for a given set of independent potential parameters has also been derived. To check the obtained results we also use a numerical method the RKGS method to solve the Schr\"odinger equation 
for both the quartic and sextic polynomial potentials with the Coulomb term. For the different sets of potential parameters the 
analytical and the numerical results are produced in tabular form and shown that our analytical results provide very good approximations. 
\begin{tabbing}
\large{\bf{Acknowledgements}}\\
\end{tabbing}
\hspace*{0.5 cm} Barnali Chakrabarti and Sudesna Bera would like to acknowledge the financial support of the DST (Govt. of India) through Contract No. SR/S2/CMP-0126/2012 dated 20/06/2014 and also WBDST through Memo No. 1211(Sanc.)/ST/P/S\&T/4G-1/2012 dated \\04/03/2016.

\end{document}